\documentclass[showpacs,preprint]{revtex4}

\usepackage{graphicx}
\usepackage{dcolumn}
\usepackage{amsmath}

\makeatletter
\def\btt#1{\texttt{\@backslashchar#1}}
\DeclareRobustCommand\bblash{\btt{\@backslashchar}} \makeatother

\begin{document}

\title[]{A nonsingular rotating black hole}
\author{Sushant~G.~Ghosh$^{a,\;b\;}$} \email{sghosh2@jmi.ac.in,
sgghosh@gmail.com} \affiliation{$^{b}$ Centre for Theoretical
Physics, Jamia Millia Islamia, New Delhi 110025, India}
\affiliation{$^{a}$ Astrophysics and Cosmology Research Unit, School
of Mathematical Sciences, University of KwaZulu-Natal, Private Bag
54001, Durban 4000, South Africa}

\date{\today}

\begin{abstract}
The spacetime singularities in classical general relativity are inevitable, which are also predicated by the celebrated singularity theorems.  However, it is general belief that singularities do not exist in the nature and they are the limitations of the general relativity.  In the absence of a well defined quantum gravity, models of regular black holes have been studied.
We employ probability distribution inspired mass function $m(r)$ to replace Kerr black hole mass $M$  to present a nonsingular rotating black hole that is identified asymptotically ($r \gg k$, $k>0$ constant) exactly as the Kerr-Newman black hole, and as the Kerr black hole when $k=0$. The radiating counterpart renders a nonsingular generalization of Carmeli's spacetime as well as Vaidya's spacetime, in the appropriate limits. The exponential correction factor changing the geometry of the classical black hole to remove curvature singularity can be alsoe motivated by the quantum arguments.  The regular rotating spacetime can also be understood as a black hole of general relativity coupled to nonlinear electrodynamics.

\end{abstract}

\pacs{04.50.Kd, 04.20.Jb, 04.40.Nr, 04.70.Bw}

\maketitle
 The celebrated theorems of Penrose and Hawking \cite{he},
state under some circumstances singularities are inevitable in
general relativity.  For the Kerr solution these singularities have
the shape of a ring, and it is timelike.  The  Kerr metric
\cite{kerr} is undoubtedly the most remarkable exact solution in the
Einstein theory of general relativity, which represents the
prototypical black hole that can arise from gravitational collapse,
which contains an event horizon \cite{bc}. Thanks to the no-hair
theorem, that the vacuum region outside a stationary black hole has
a Kerr geometry.  It is believed that spacetime singularities do not
exist in Nature; they  are creation of the general relativity.  It
turns out that what amount to a singularity in the general relativity
could be adequately explained by some other  theory, say, quantum
gravity. However, we are yet afar from a specific theory of quantum
gravity. So a suitable course of action is to understand the inside
of a black hole and resolve its singularity by carrying out research
of classical black holes, with regular (nonsingular) properties,
where spacetime singularities can be avoided in presence of
horizons. Thus, the regular black holes are solutions that have horizons and, contrary to classical black holes which have singularities at the
origin, their metrics as well as their curvature invariants are regular everywhere \cite{sa}. This can be motivated by quantum arguments of  Sakharov
\cite{ads} and Gliner \cite{Gliner} who proposed that spacetime in
the highly dense central region of a black hole should be de
Sitter-like for $r \simeq 0$, which was later explored and refined
by Mukhanov and his collaborators \cite{Mukhanov}.  This indicates
that an unlimited increase of spacetime curvature during a collapse
process, which may halt it, if quantum fluctuations dominate the
process. This puts an upper bound on the value of curvature and
compels the formation of a central core.

Bardeen \cite{Bardeen} realized  the idea of a central matter core,
by proposing the first regular black hole, replacing the singularity
by a regular de Sitter core, which is solution of the Einstein
equations coupled to an electromagnetic field, yielding a alteration of
the Reissner-Nordstr$\ddot{o}$m metric. However, the
physical source associated to Bardeen  solution was clarified much later by Ayon-Beato and
Garcia \cite{ABG99}.   The exact
self-consistent solutions for the regular black hole for the
dynamics of gravity coupled to nonlinear electrodynamics have been also
obtained later \cite{AGB}, which also shares most properties of the
Bardeen's black hole. Subsequently, there has been intense
activities in the investigation of regular black holes
\cite{regular,sa,Hayward}, and more recently \cite{Xiang,hc,lbev,Balart:2014cga}, but most of these solutions are more or less based on Bardeen's
proposal.
  However, non-rotating black holes cannot be tested
by astrophysical observations, as the black hole spin plays  a
critical and key role in any astrophysical process. This prompted
generalization of these regular solutions to the axially symmetric
case or to the Kerr-like solution \cite{Bambi,Neves:2014aba,Toshmatov:2014nya,Ghosh:2014hea,Larranaga:2014uca}, via the Newman-Janis algorithm
\cite{nja} and by other similar techniques \cite{Azreg-Ainou:2014aqa,Azreg-Ainou:2014nra,Azreg-Ainou:2014pra}.  It is also demonstrated that these rotating regular solutions can act as particle accelerator \cite{Amir:2015pja,Ghosh:2014mea}.   The algorithm also allowed to generate
a noncommutative inspired rotating black holes with a regular
de Sitter toroidal region \cite{Modesto:2010rv}. However, these
rotating regular solutions go over to the Kerr solution in the appropriate limits, but
not to the Kerr-Newman solution.

\begin{figure*}
    \begin{tabular}{c c}
        \includegraphics[scale=0.55]{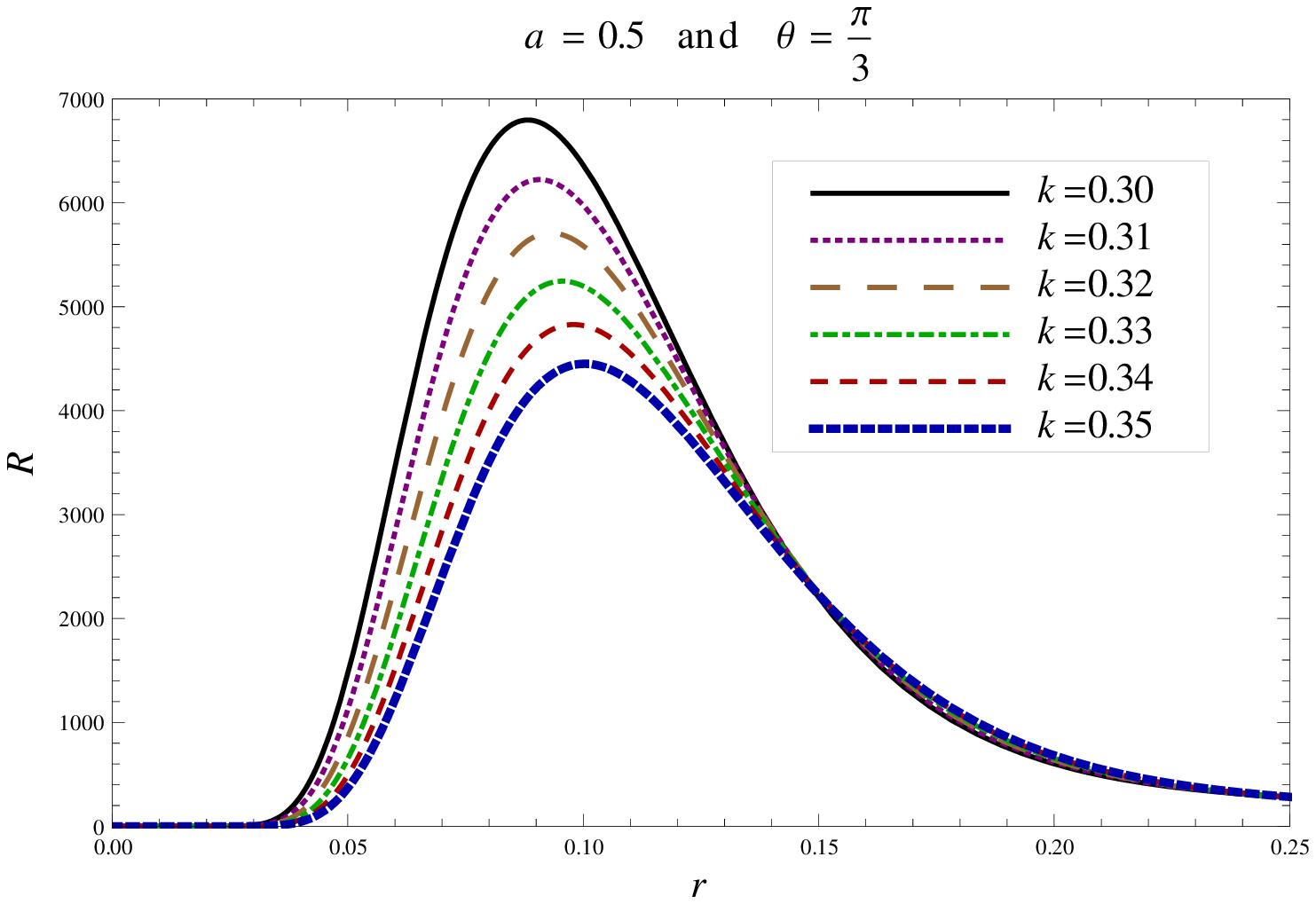}&
        \includegraphics[scale=0.55]{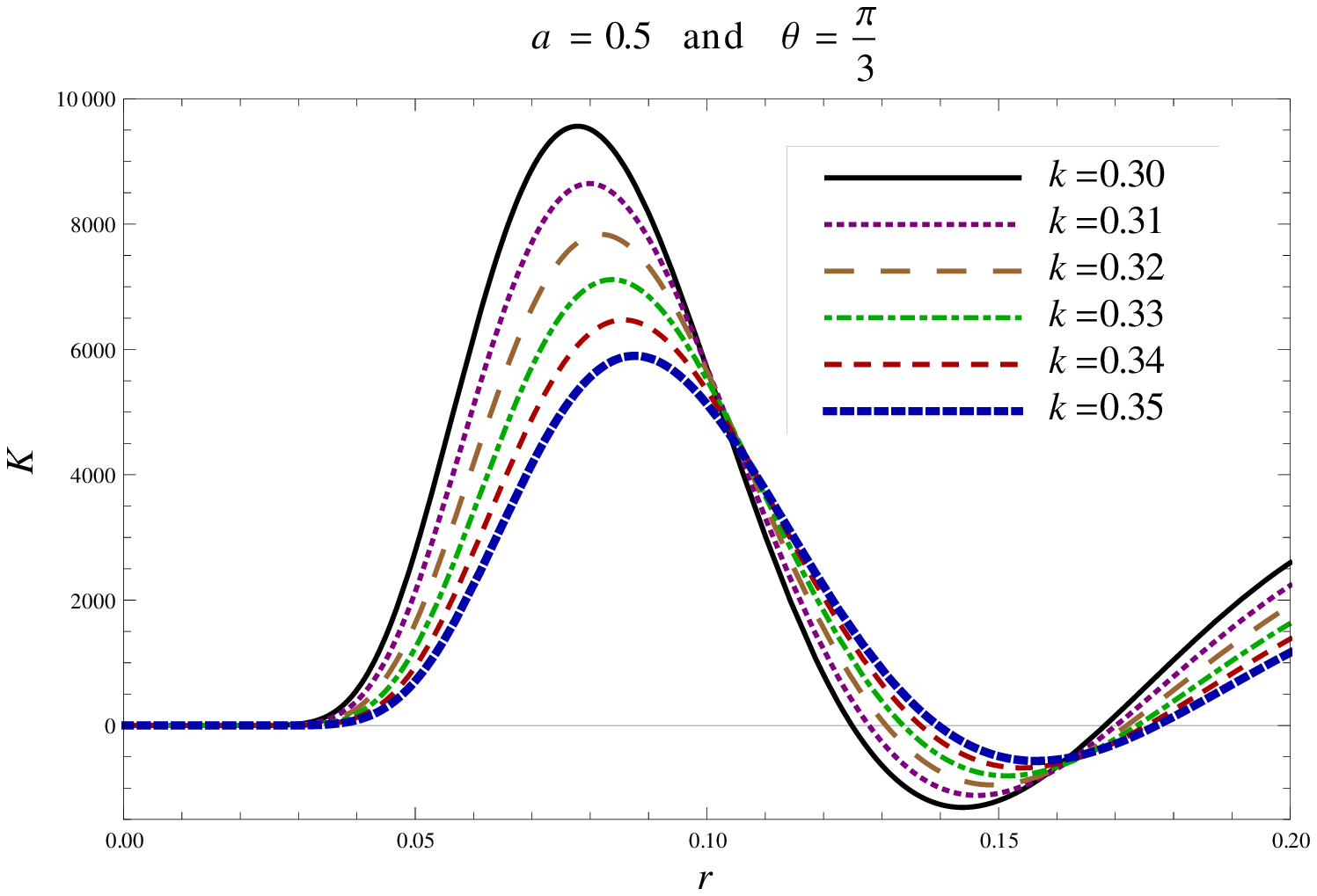}\\
         \includegraphics[scale=0.6]{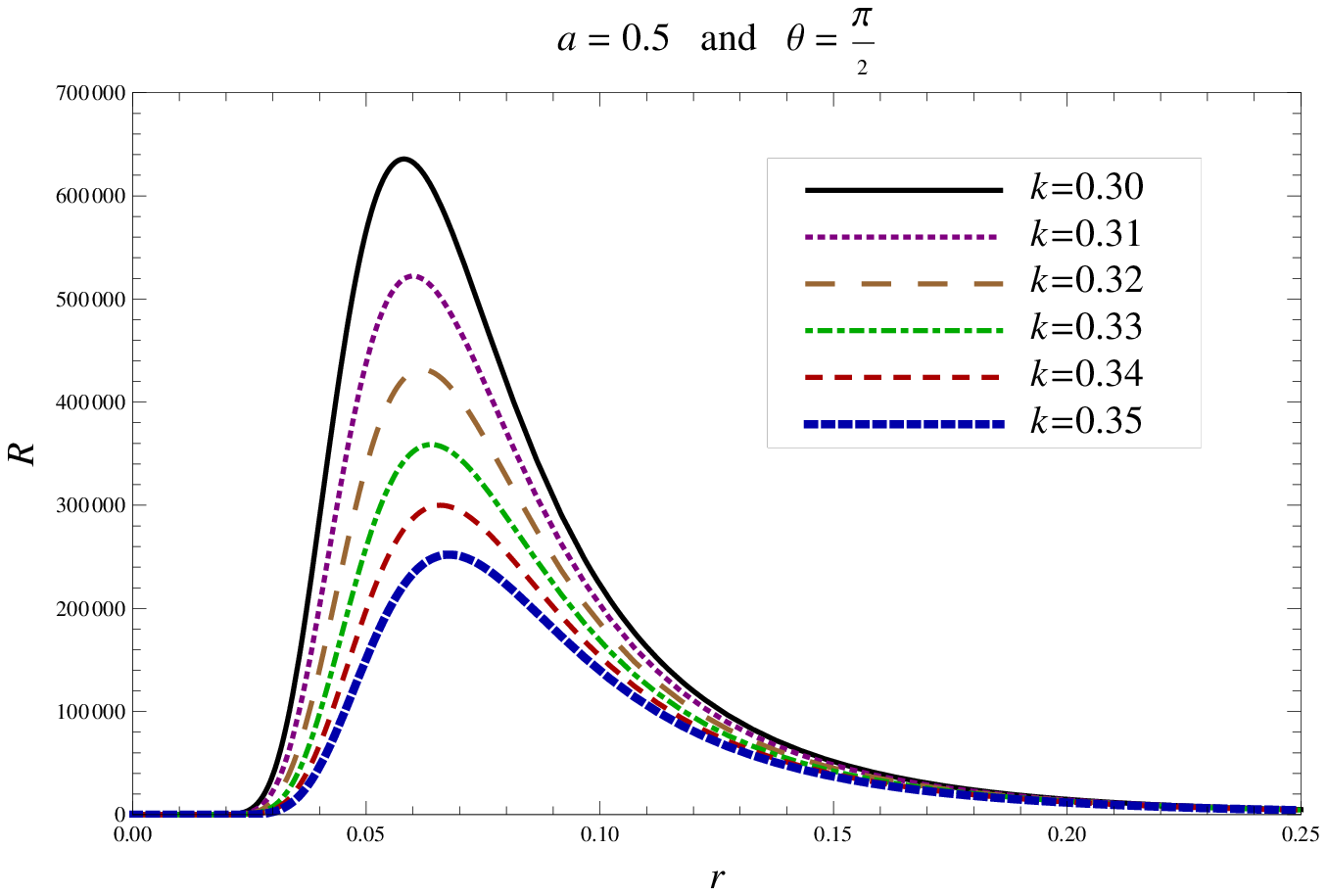}&
                \includegraphics[scale=0.6]{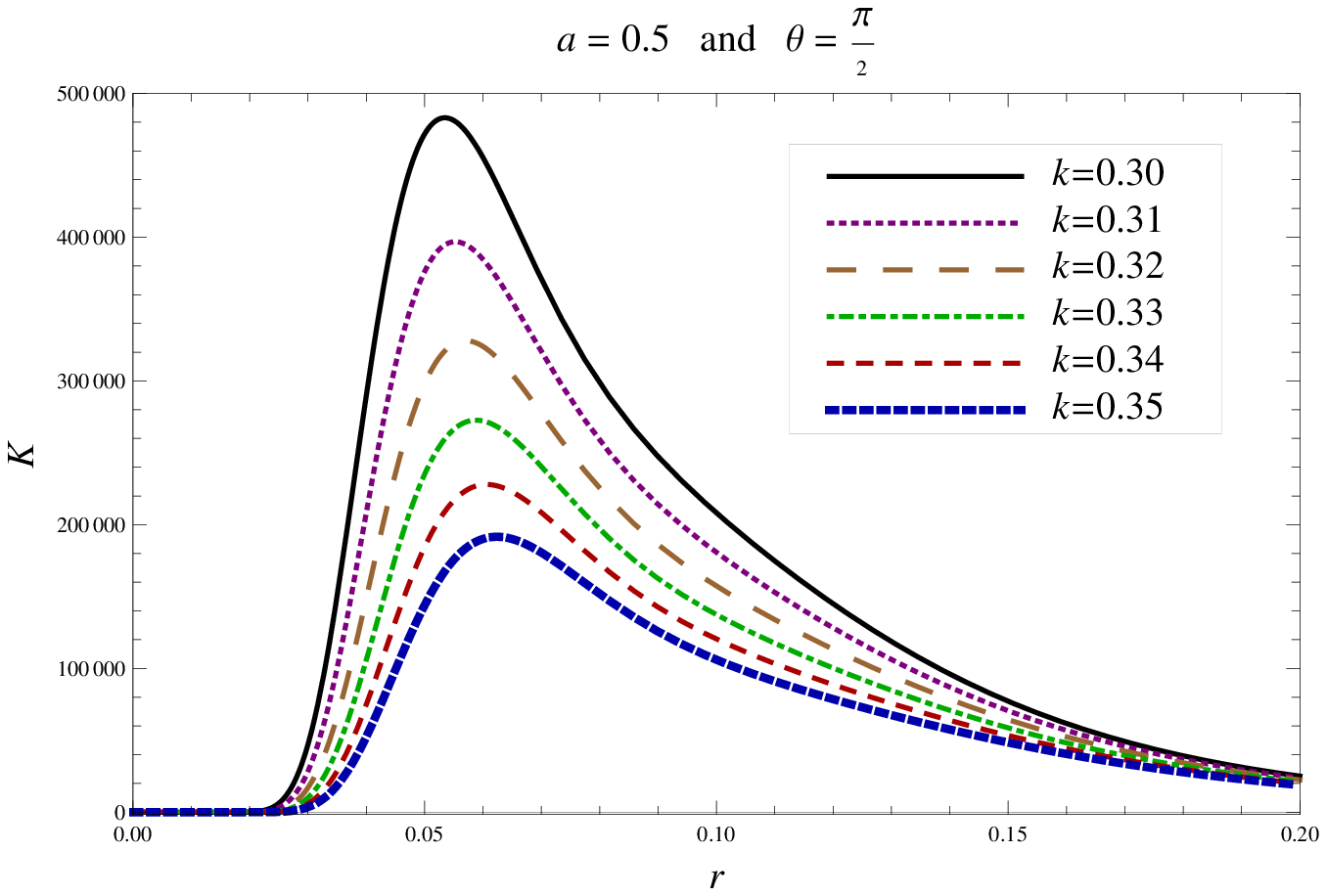}
    \end{tabular}
    \caption{\label{ks}Plots showing the regular behavior of the Ricci
        scalar and Kretschmann scalar vs radius for the different values of
        parameter $k$ (with $M=1$).}
\end{figure*}
This Letter searches for a new class of three parameter stationary, axisymmetric
metrics that describe a regular (nonsingular) rotating black holes.
The metrics depend on the mass ($M$) and spin ($a$) as well a free parameter $(k)$ that measure potential
deviation from the Kerr solution \cite{kerr} and also generalizes the Kerr-Newman solution \cite{nja}, which in the
Boyer-Lindquist coordinates reads
\begin{eqnarray}\label{rotbhtr}
ds^2 & = & - \left( 1- \frac{2Mr e^{-k/r}}{\Sigma} \right) dt^2 +
\frac{\Sigma}{\Delta}dr^2 + \Sigma d \theta^2\nonumber
\\ & - & \frac{4aMr
e^{-k/r}}{\Sigma  } \sin^2 \theta dt \; d\phi \nonumber
\\ & + & \left[r^2+ a^2 +
 \frac{2M r a^2 e^{-k/r}}{\Sigma} \sin^2 \theta
\right] \sin^2 \theta d\phi^2,
\end{eqnarray}
with $\Sigma = r^2 + a^2 \cos^2\theta$, $\Delta=r^2 + a^2 - 2 M r
e^{-k/r}$, and $M,\;a,$ and $\;k$ are three parameters, which will
be assumed to be positive. The metric (\ref{rotbhtr}) includes the Kerr
solution as the special case if deviation parameter, $k=0$, and
 the Schwarzschild solution for $k=a=0$.  In that case $M=0$, the  metric (\ref{rotbhtr}) actually is
 nothing more than  the Minkowski spacetime expressed in spheroidal
 coordinates.  When only $a=0$,  the rotating regular metrics (\ref{rotbhtr}) to
 \begin{equation}\label{schw}
 - \left(1 - \frac{2Me^{-k/r}}{r}\right) dt^2+ \left(1 - \frac{2Me^{-k/r}}{r}\right)^{-1} dr^2 + r^2 d\Omega^2
 \end{equation}
 with $d \Omega^2= d\theta^2+\sin^2\theta d \phi^2.$ It is a spherically symmetrical regular metrics \cite{hc,Balart:2014cga,Xiang}, which is a modification of Schwarzschild solution.  It is easy to see that by employing the Newman-Janis algorithm \cite{nja} to static spherical regular solution (\ref{schw}), one obtains the rotating regular spacetime (\ref{rotbhtr}).  The properties of metric (\ref{schw}) including the thermodynamics has been also analyzed \cite{Xiang,hc}
 Interestingly, the exponential  convergence factor is used in the formulation
 of the quantum gravity that is finite to all order in the Planck length \cite{Brown}. Further, the inclusion of such quantum  gravity effects
 makes other flat space quantum field theories similarly finite. Further, A finite quantum gravity theory can be used to resolve the cosmological
constant problem \cite{Moffat}. Thus, the rotating regular metrics (\ref{rotbhtr}) is same as the Kerr black hole, but the mass ($ M $) of the Kerr black hole is changed to $m(r)$ as
\begin{equation}\label{massr}
m(r) = \frac{\sigma(r)}{\sigma_{\infty}} M,
\end{equation}
where the function $\sigma(r) \geq 0$ and $\sigma'(r) < 0$ for $r\geq0$, and $\sigma(r)/r \rightarrow 0$ as $r\rightarrow0$, and  $\sigma(\infty) = \sigma(r \rightarrow \infty)$ denotes the normalization factor.  Here,  $\sigma(r) = \exp(-k/r)$ so that  $\sigma(\infty) = 1$.

Note that the metric (\ref{rotbhtr}) asymptotically ($r \gg k$)
behaves as rotating counterpart of the Reissner-Nordstr$\ddot{o}$m
solution or the Kerr-Newman solution \cite{nja}, i.e.,
\begin{eqnarray*}
  g_{tt} &=& 1 - \frac{(2Mr-q^2)}{\Sigma} + \mathcal{O}(k^2/r^2), \\
  \Delta &=& r^2 + a^2 - 2 M r + q^2 + \mathcal{O}(k^2/r^2).
\end{eqnarray*}
This happens when the charge $q$ and mass $M$ are related to the
parameter $k$ via $q^2=2Mk$.  Thus, the solution (\ref{rotbhtr}) is
stationary, axisymmetric with killing field
$(\frac{\partial}{\partial t})^a$ and $(\frac{\partial}{\partial
\phi})^a$, and  all known stationary black holes are
encompassed by the three parameter family solutions, and it
also generalizes the Kerr-Newman solution\cite{nja}.
\begin{figure*}
    \begin{tabular}{c c}
        \includegraphics[scale=0.6]{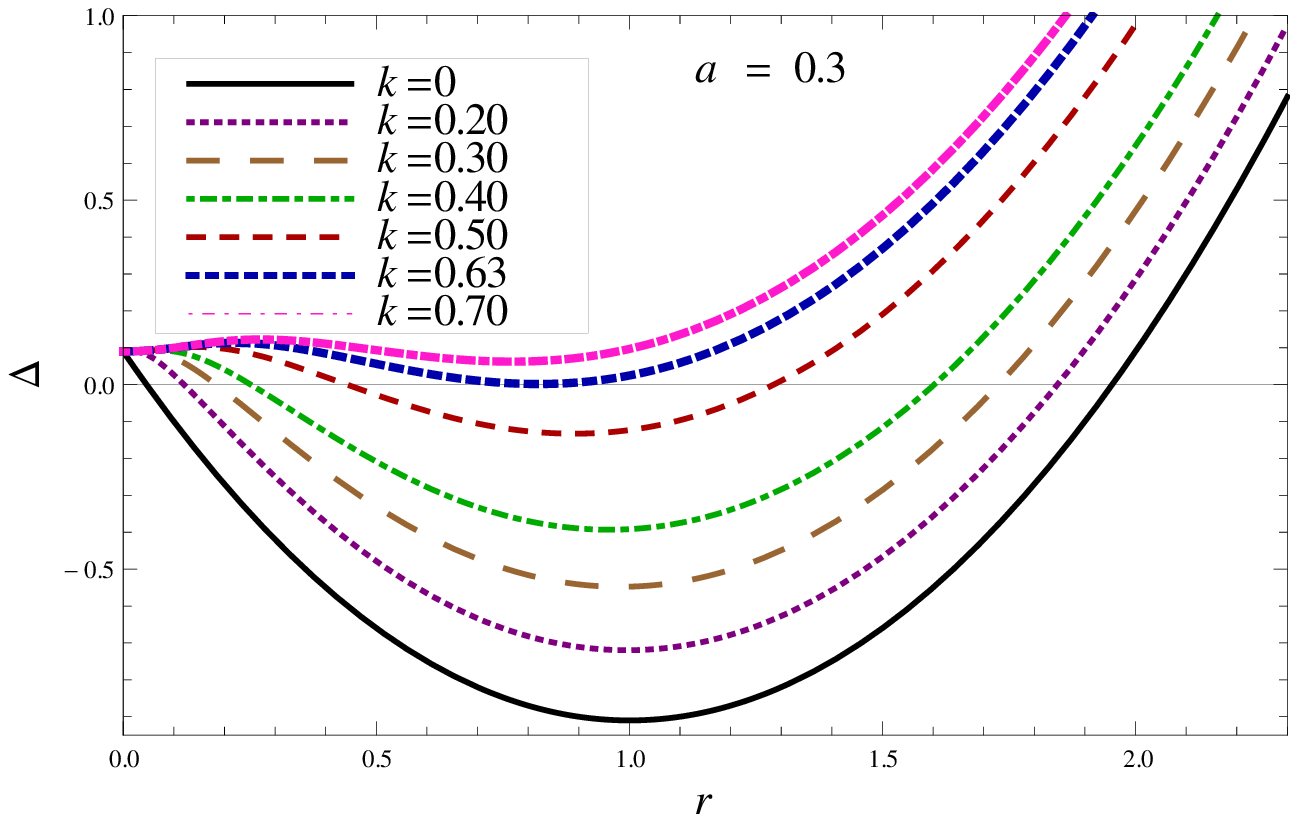}&
        \includegraphics[scale=0.6]{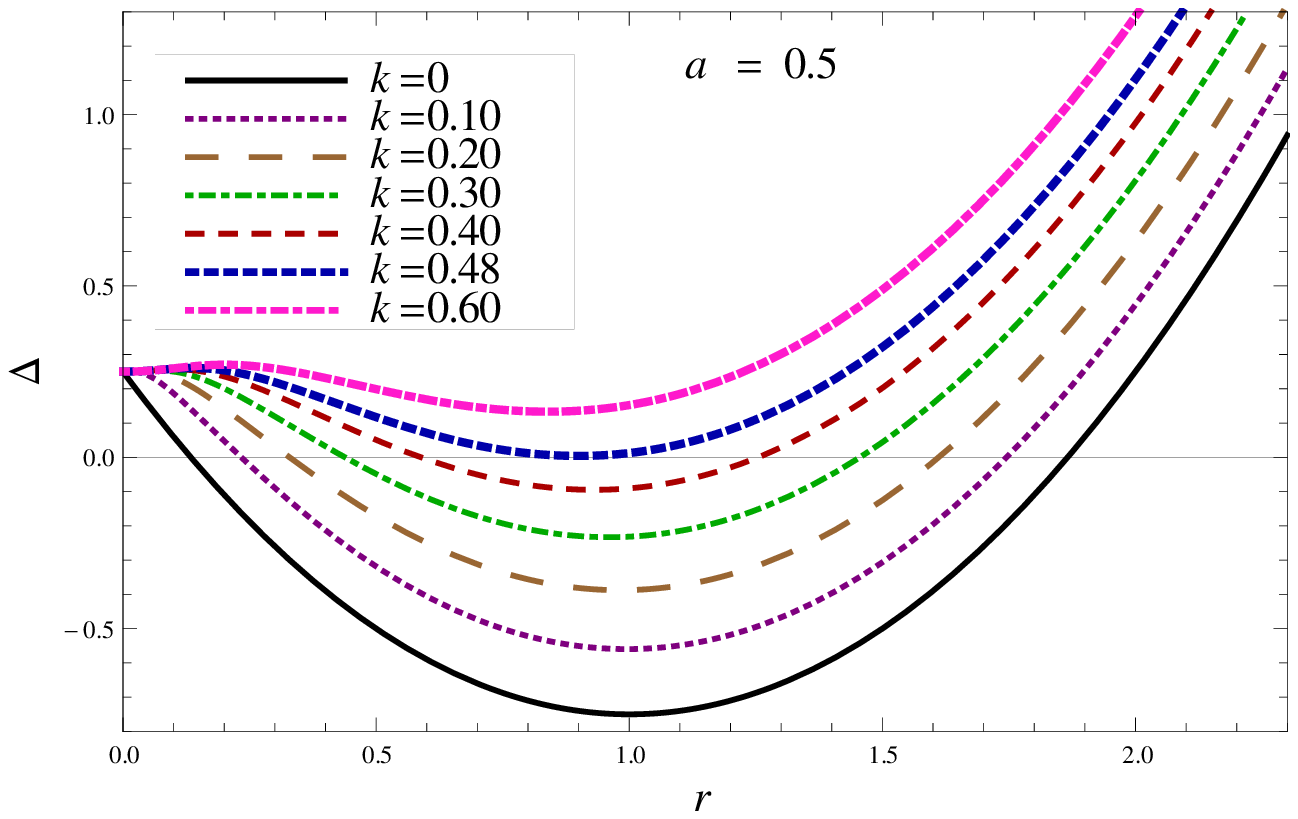}
    \end{tabular}
    \caption{Plots showing the behavior of $\Delta$ vs radius for
        different values of parameter $k$ (with $M=1$); Case $k=0$ corresponds to Kerr black hole.}\label{ehf}
\end{figure*}
It is not difficult to find numerically a range of $M$ and $k$ for
which the solution (\ref{rotbhtr}) is a black hole, in addition to
regular everywhere. Henceforth, for definiteness we shall address
the solution (\ref{rotbhtr}) as the regular (nonsingular) rotating
black hole.

We approach regularity problem of solution by studying the behavior
of  invariant $\mbox{R} = R_{ab} R^{ab}$ ($R_{ab}$ is the
Ricci tensor) and the Kretschmann invariant $\mbox{K} = R_{abcd}
R^{abcd}$ ($R_{abcd}$ is the Riemann tensor). It may be pointed out that the solution (\ref{rotbhtr}) is regular if the  curvature invariants are well behaved \cite{he,Bambi,AGB,Neves:2014aba}
(\ref{rotbhtr}), they read
\begin{eqnarray}
R &=&
\frac{2k^{2}M^{2}e^{\frac{-2k}{r}}}{r^{6}\Sigma^{4}}\left(\Sigma^{2}k^{2}-4r^{3}\Sigma
k + 8r^{6}\right) ,\nonumber\\
K & =
&\frac{4M^{2}e^{\frac{-2k}{r}}}{r^{6}\Sigma^{6}}\left(\Sigma^{4}k^{4}-8r^{3}\Sigma^{3}k^{3}
 +A k^{2}+B k+C\right),
\end{eqnarray}
where $A, B$ and $C$ are functions of $r$ and $\theta$ given by
\begin{eqnarray*}
A &=& -24 r^{4}\Sigma (-r^{4}+a^{4}\cos^{4}\theta),\nonumber\\
B &=& -24 r^{5}\left(r^{6}+a^{6}\cos^{6}\theta-5r^{2}a^{2}\cos^{2}\theta\Sigma\right),\nonumber\\
C &=& 12
r^{6}\left(r^{6}-a^{6}\cos^{6}\theta\right)-180r^{8}a^{2}\cos^{2}\theta\left(r^{2}-a^{2}\cos^{2}\theta\right).
\end{eqnarray*}
These invariants, for $M \neq0$, are regular everywhere, including
at $\Sigma=0$, where noteworthy they vanish (cf.
Fig~\ref{ks}).

\begin{table}
    \begin{center}
        \caption{Radius of EHs, SLSs and    $\delta^{a}=r^{+}_{SLS}-r^{+}_{EH}$ for different values of
            parameter $k$ (with $M=1$ and $\theta=\pi/3$).}\label{tes}
        \begin{tabular}{l l l l | l l l }
            \hline \hline
            &\multicolumn{3}{c}{$a=0.3$}  &  \multicolumn{3}{c}{$a=0.5$}\\
            \hline
            $k$ & $r^{+}_{EH}$ & $r^{+}_{SLS}$ & $\delta^{0.3}$ & $r^{+}_{EH}$ & $r^{+}_{SLS}$ & $\delta^{0.5}$   \\
            \hline
            0   & 1.95394      & 1.98869       & 0.03475  & 1.86603      & 1.96825       & 0.10222    \\
            0.1 & 1.84577      & 1.88471       & 0.03894  & 1.74540      & 1.86184       & 0.11644    \\
            0.2 & 1.72956      & 1.77409       & 0.04453  & 1.61141      & 1.74802       & 0.13661    \\
            0.3 & 1.60235      & 1.65479       & 0.05244  & 1.45583      & 1.62422       & 0.16839    \\
            0.4 & 1.45860      & 1.52336       & 0.06476  & 1.25489      & 1.48593       & 0.23104    \\
            0.45& 1.37716      & 1.45129       & 0.07413  & 1.10409      & 1.40876       & 0.30467    \\
            \hline \hline
        \end{tabular}
    \end{center}
\end{table}

We have also examined the other Ricci and Weyl invariant for the rotating metric (\ref{rotbhtr}) \cite{invar}, they can be obtained with Mathematica or Maple.  It will be necessary to introduce the trace-free Ricci tensor defined by $S^a_b=R^a_b-\delta^b_a R/4$, where $R^a_b$ is Ricci tensor \cite{invar}.  The non-zero Ricci invariant for the rotating metric (\ref{rotbhtr}) reads
\begin{eqnarray}
&& RS=R^a_a= \frac{2k^2 M e^{-k/r}}{r^3\Sigma} \nonumber \\
&& R1= \frac{1}{4} S^b_a \; S^a_b =  \frac{k^2 M^2 e^{-2k/r} (k\Sigma-4r^3)^2 }{ r^6 \Sigma^4} \nonumber \\
&& R3= \frac{1}{16} S^b_a \; S^c_b\; S^d_c \; S^a_d = \frac{1}{64} \frac{k^4 M^4 e^{-4k/r} (k\Sigma-4r^3)^4}{r^12\Sigma^8 }.
\end{eqnarray}
Not surprisingly which are also regular everywhere. The Weyl invariant $W1I$ reads:
\begin{align}
W1I = \frac{1}{8} = C_{ab}^{*cd} C_{cd}^{ab} = \frac{2a M^2 \cos\theta e^{-2k/r} \alpha \beta}{r^4\Sigma^6}
\end{align}
with
\begin{eqnarray}
& & \alpha= {k}^{2}{a}^{4}  \cos^{4}\theta +2\,{r}^
{2}{a}^{2} \left( -9\,{r}^{2}-3\,rk+{k}^{2} \right)   \cos^{2}\theta +{r}^{4} \left( {k}^{2}+6\,{r}^{2}
-6\,rk \right)  \nonumber \\
& & \beta= {a}^{2} \left( r+k \right)   \cos^{2}\theta +{r}^{2} \left( -3\,r+k \right).
\end{eqnarray}
Here $C_{abcd}$ is the Weyl (conformal) tensor, and $C_{abcd}^{*}$ its tensor dual.
We shall not report the analytic form of other Weyl invariant which are also obtained and found to be regular everywhere.  Thus, the exponential factor $e^{-k/r}$ removes the curvature singularity of the  Kerr black hole.

In addition, the solution (\ref{rotbhtr}) is singular at the points
where $\Sigma \neq 0$, and $\Delta=0$, is a coordinate singularity
this surface is called event horizon (EH). The numerical analysis of
the transcendental equation $\Delta=0$ reveals that it is possible
to find non-vanishing values of parameters $a$ and $k$ for which
$\Delta$ has a minimum, and it admits two positive roots $r_{\pm}$.
It turns out that  $r=r_{\pm}$ are coordinate singularities of the
same nature as the singularity at $r=2M$ in the Schwarzschild
spacetime, the metric can be smoothly extended across $ r = r_+ $,
with {$ r = r_+ $} being a smooth null hypersurface, and the
simplest possible extension could be rewriting solution
(\ref{rotbhtr}) in Eddington-Finkelstein coordinates as shown below
(cf. Eq.~ (\ref{rotbhvr})).   It turns out that, for a given $a$,
there exists a critical value of $k$, $k^{EH}_c$, and one of $r$,
$r^{EH}_c$, such that $\Delta=0$ has a double root which corresponds
to a regular extremal black hole with degenerate horizons
($r^{EH}_-=r^{EH}_+=r^{EH}_c$). When $k<k^{EH}_c$, $\Delta=0$ has
two simple zeros, and has no zeros for $k>k^{EH}_c$ (cf.
Fig.~\ref{ehf}). These two cases corresponds, respectively, a
regular non-extremal black hole with a Cauchy horizon and an EH, and
a regular spacetime.  It is worthwhile to mention that the critical
values of $k^{EH}_c$ and $r_c^{EH}$ are $a$ dependent, e.g., for
$a=0.3,\;0.5$, respectively $k^{EH}_c=0.63,0.48$ and
$r_c^{EH}=0.82,0.89$ (cf. Fig.~\ref{ehf}). Indeed, the $k^{EH}_c$
decreases with the increase in $a$, on the other hand, the radius
$r_c^{EH}$ increases with an increase in $a$.
\begin{figure*}
\begin{tabular}{c c}
\includegraphics[scale=0.6]{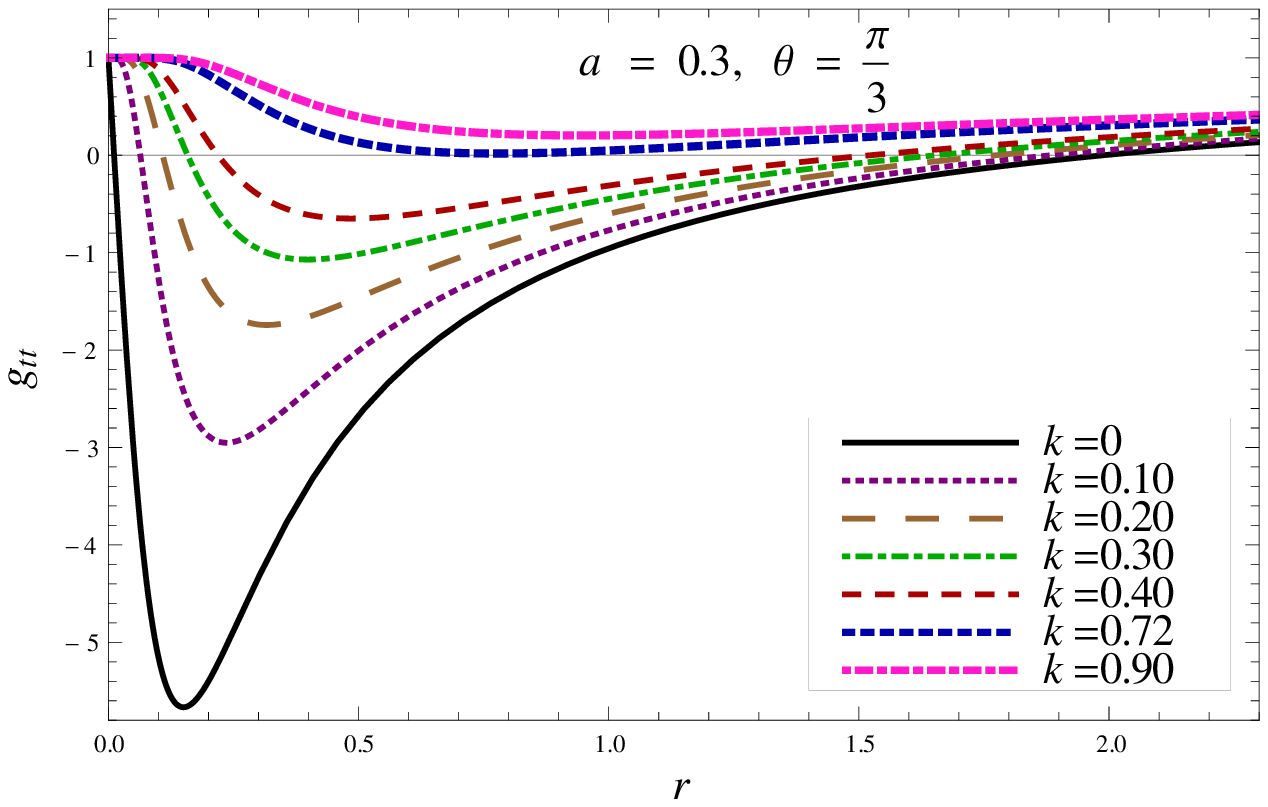}&
\includegraphics[scale=0.6]{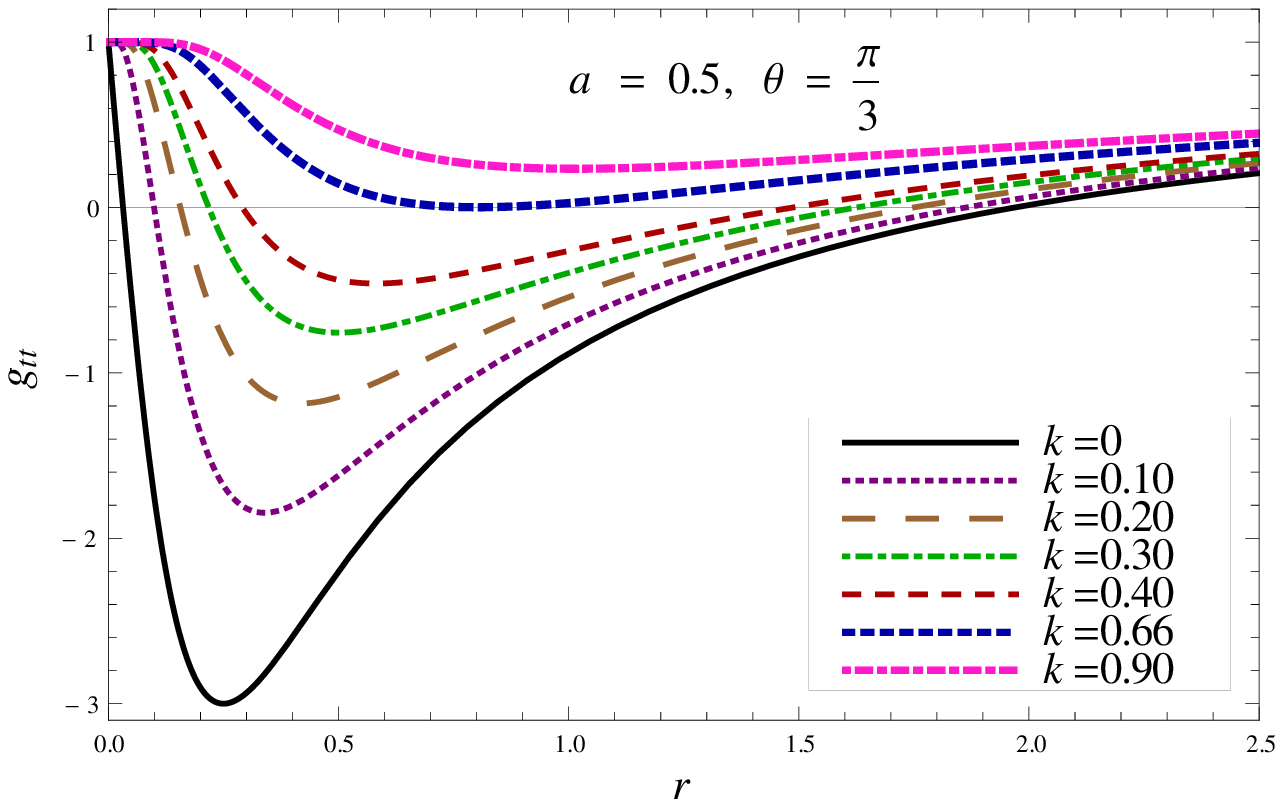}
\end{tabular}
\caption{Plots showing the behavior of -$g_{tt}$ vs radius for the
different values of parameter $k$  (with $M=1$); Case $k=0$ corresponds to Kerr
black hole.}\label{slsf}
\end{figure*}
The timelike killing vector $\xi^a =(\frac{\partial}{\partial t})^a$
of the solution has norm
\begin{equation}\label{tls}
\xi^a \xi_a = g_{tt} = - \left( 1- \frac{2Mr e^{-k/r}}{\Sigma
}\right),
\end{equation}
becomes positive in the region where $ r^2 + a^2 \cos^2 \theta - 2 M e^{-k/r} <0. $ This killing vector is null at the stationary limit
surface (SLS), whose locations are, for different $k$,  depicted in
Fig.\ref{slsf}. The analysis of the zeros of the $g_{tt}=0$ , for a
given value of $a$ and $\theta$, disseminate a critical parameter
$k_c^{SLS}$ such that $g_{tt}=0$ (e.g., $k_c^{SLS}=0.72\;
\mbox{and}\; 0.66$, respectively for $a=0.3\; \mbox{and}\; 0.5 $)
has no roots if $k>k_c^{SLS}$, a double root at $k=k_c^{SLS}$, two
simple zeros if $k<k_c^{SLS}$ (cf. Fig~\ref{slsf}). Also, like in
the case of EH, $k_c^{SLS}$ and $r_c^{SLS}$ also have similar
behavior with $a$ for a given $\theta$ as shown in Fig.~\ref{slsf}.
Interestingly, the radii, of EH and SLS for the solution decreases
when compared to analogous Kerr case ($k=0$). Notice, that for
$\theta=0,\;\pi$, the SLS and EH coincides.  On other hand, outside
this symmetry, they do not (cf. Table \ref{tes}) as in the usual
Kerr/Kerr-Newman. The region between $r_+^{EH}\, < r\, < r_+^{SLS}$
is called \textit{ergosphere}, where the asymptotic time translation
Killing field $\xi^a=(\frac{\partial}{\partial t})^a$ becomes
spacelike and an observer follow an orbit of $\xi^a$. The shape of
the ergosphere, therefore, depends on the spin $a$, and parameter
$k$.  It came as a great surprise when Penrose \cite{pc}
 suggested that energy can be extracted from a black hole with an ergosphere.
 On the other hand, the Penrose process \cite{pc}  relies on
the presence of an \textit{ergosphere},   which for the solution
(\ref{rotbhtr}) grows with the increase of parameter $k$ as well
with spin $a$ as demonstrated in Table \ref{tes}. This in turn is
likely to have impact on energy extraction, which is being
investigated separately. The vacuum state is obtained by letting
horizons size go to zero or by making black hole disappear this
amounts to $r \rightarrow \infty$.  One thus conclude that the
solution is asymptotically flat as the metric components approaches
those  of the Minkowski  spacetime in spheroidal coordinates.

In order to further analyze the matter associated with the metric (\ref{rotbhtr}), we use orthonormal basis in which energy-momentum tensor is diagonal  \cite{Bambi,Neves:2014aba,bpt}
\begin{equation}
e^{(a)}_{\mu}=\left(\begin{array}{cccc}
\sqrt{\mp(g_{tt}-\Omega g_{t\phi})}& 0 & 0 & 0\\
0 & \sqrt{\pm g_{rr}} & 0 & 0\\
0 & 0 & \sqrt{g_{\theta \theta}} & 0\\
{g_{t\phi}}/{\sqrt{g_{\phi\phi}}} & 0 & 0 & \sqrt{g_{\phi\phi}}
\end{array}\right),\label{Matrix}
\end{equation}
 with $\Omega= g_{t\phi} /{g_{\phi\phi}}$.  
Clearly, the metric is regular at the centre. The components of the energy-momentum tensor in the orthonormal frame reads
\begin{equation}
T^{(a)(b)} = e^{(a)}_{\mu} e^{(b)}_{\nu} G^{\mu \nu}. \nonumber
 \end{equation}
Considering the line element (\ref{rotbhtr}),
 we can write the components of the respective energy momentum
 tensor as
 \begin{eqnarray}
 \rho = \frac{2 M k e^{-k/r}}{\Sigma^2} = -P_1 \nonumber \\
 P_2 =  - \frac{M k e^{-k/r} \left(k \Sigma - 2 r^3\right)}{r^3 \Sigma^2} = P_3
 \end{eqnarray}
To check the weak energy condition, we can choose an appropriate
 orthonormal basis \cite{Bambi,Neves:2014aba,bpt} in which the energy momentum tensor reads
 \begin{equation}
 T^{(a)(b)} = \mbox{diag}(\rho, P_1,P_2,P_3)
 \end{equation}
 These stresses vanish when  $k=0$,  also for $M=0$,  fall off
 rapidly at large $r$ for $M,k \neq 0$, and for $r \gg k$ and $a=0$,
 they are, to $O(k^2/r^2)$, exactly stress energy tensor of Maxwell
 charge given by
 \[
T^{(a)}_{(b)} = T^a_b= \frac{q^2}{r^4} {\mbox{diag}}[-1, -1, 1, 1].
  \]
  In this
 limit, the solution exactly takes form of the Kerr-Newman solution.
 Further, the causal (horizon) structure of the solution
 (\ref{rotbhtr}) is similar to that of the Kerr solutions,
 except that the scalar polynomial singularity of the
 the Kerr solution, at center ($r=0$), is no more exists
 with regular behaviors of the scalars at the center as shown in
 Fig.~\ref{ks}).  Thus, the solution (\ref{rotbhtr}), which
 asymptotically behaves as the Kerr-Newman,  can be understood as a
 rotating regular black hole of general relativity coupled to a
 suitable \textit{nonlinear electrodynamics}.

\begin{figure*}
\begin{tabular}{c c}
\includegraphics[scale=0.7]{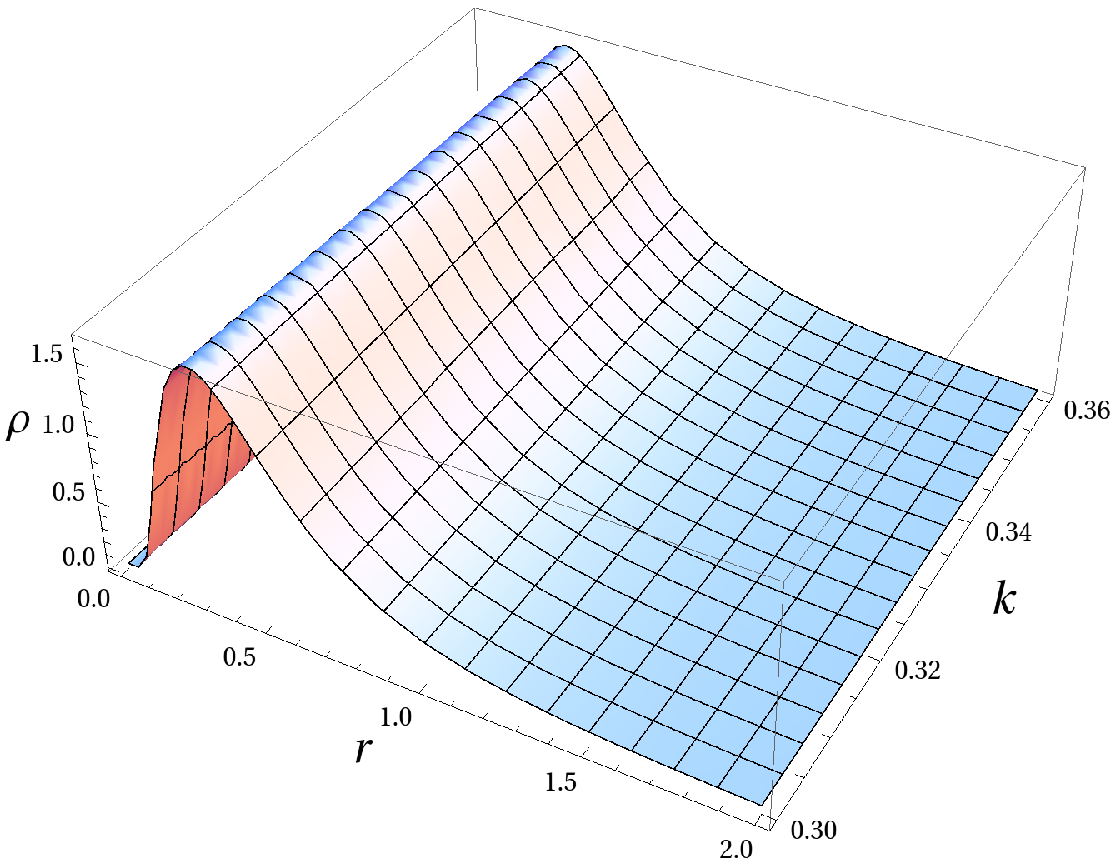} &
\includegraphics[scale=0.7]{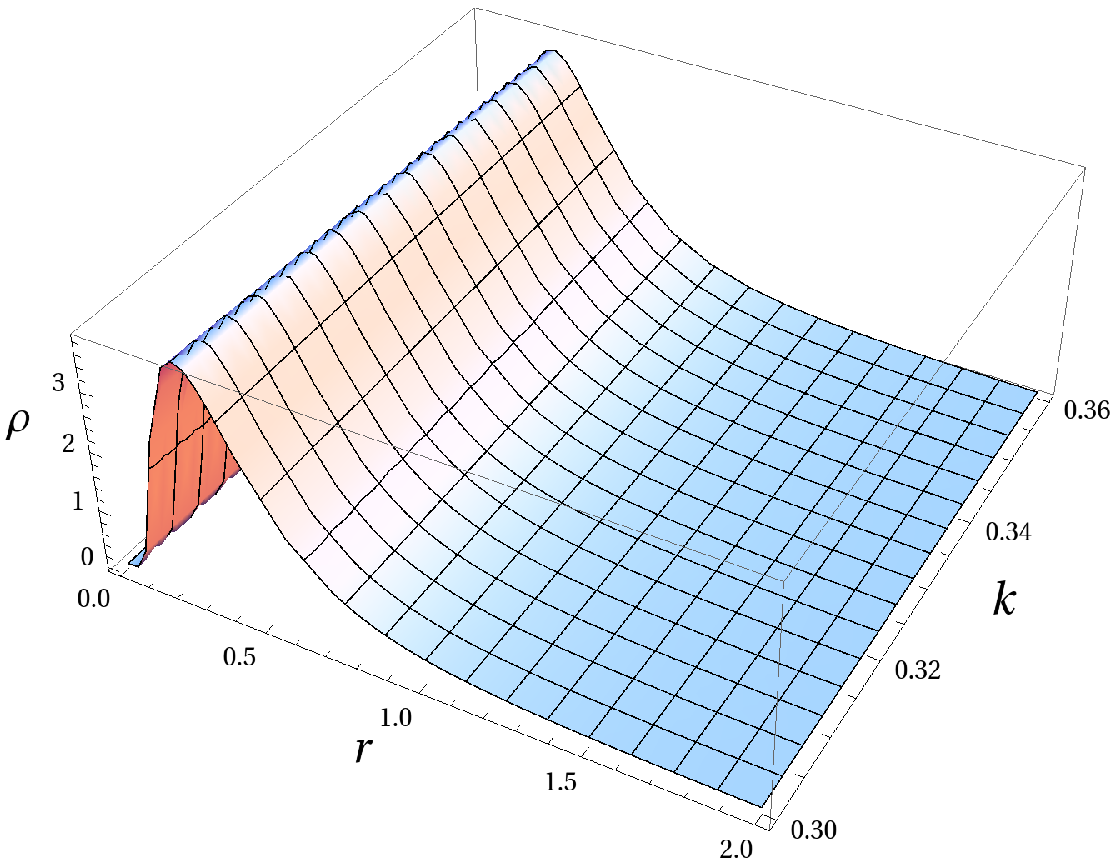} \\
\includegraphics[scale=0.7]{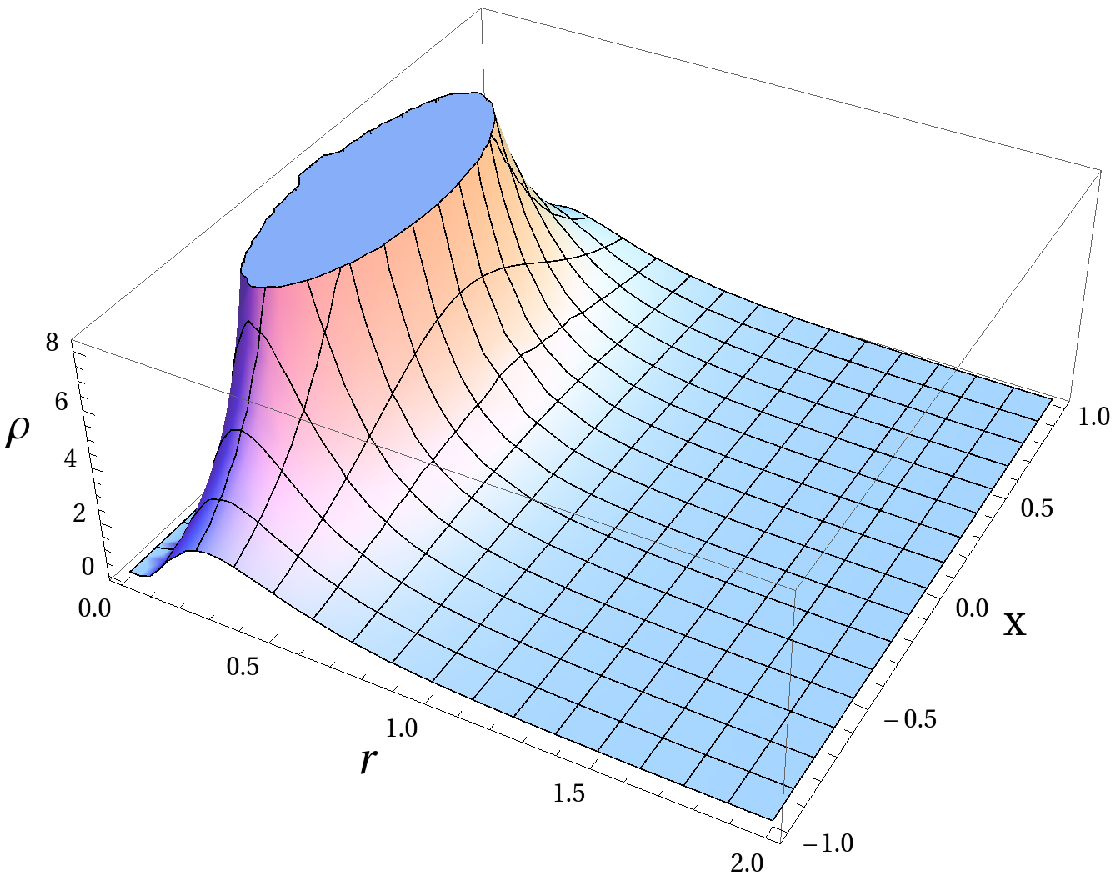} &
\includegraphics[scale=0.7]{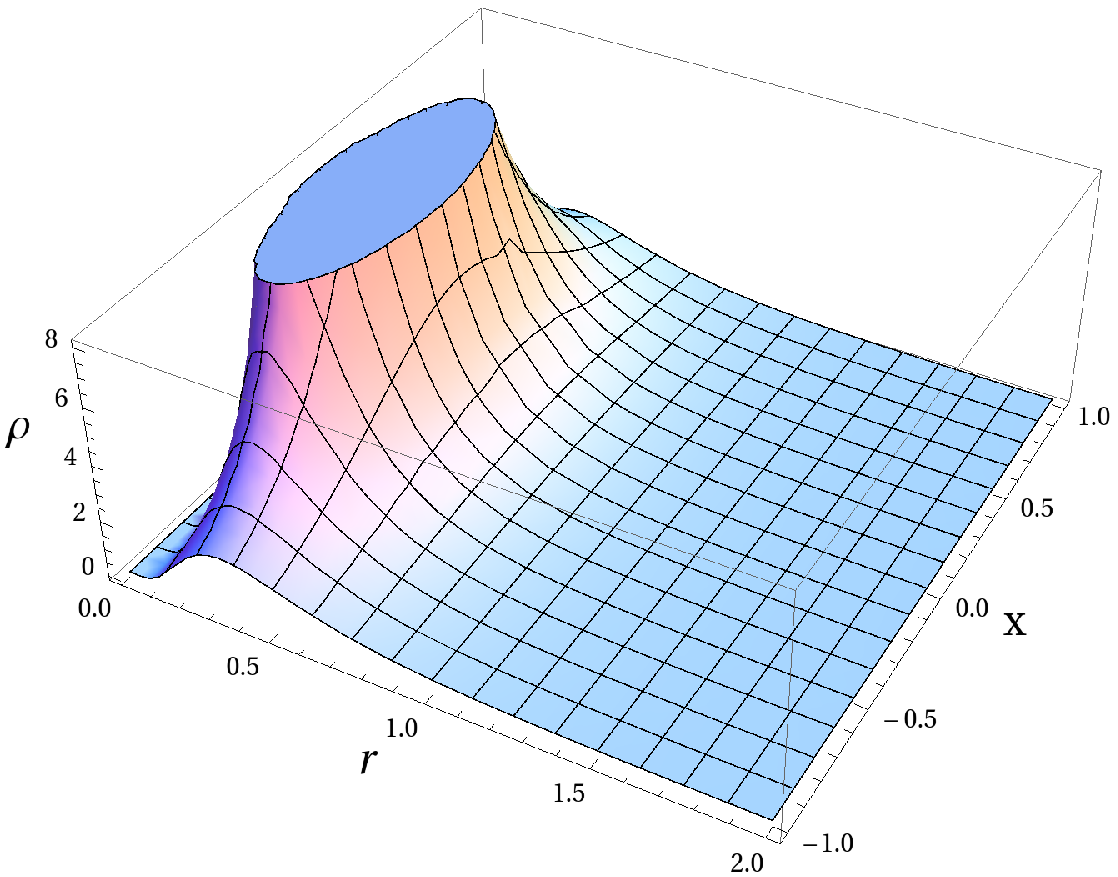}
\end{tabular}
\caption{Plots of $\rho$ vs radius (a) top for the
different values of parameter $k$ ($\theta=\pi/3$ (left) $\pi/4$ (right) ) and (b) bottom for different values of $x= \cos \theta $ ($k=0.3$ (left) $k=0.4$ (right) ) }; \label{rho}
\end{figure*}

\begin{figure*}
\begin{tabular}{c c}
\includegraphics[scale=0.7]{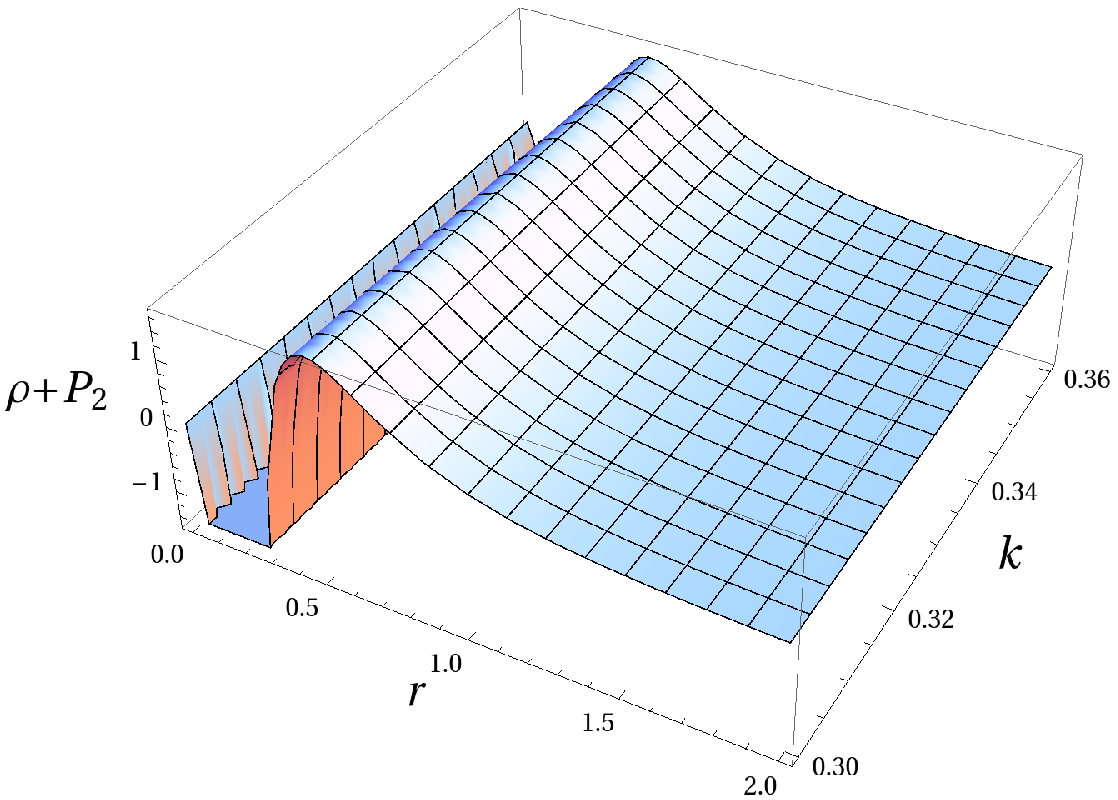}  &
\includegraphics[scale=0.7]{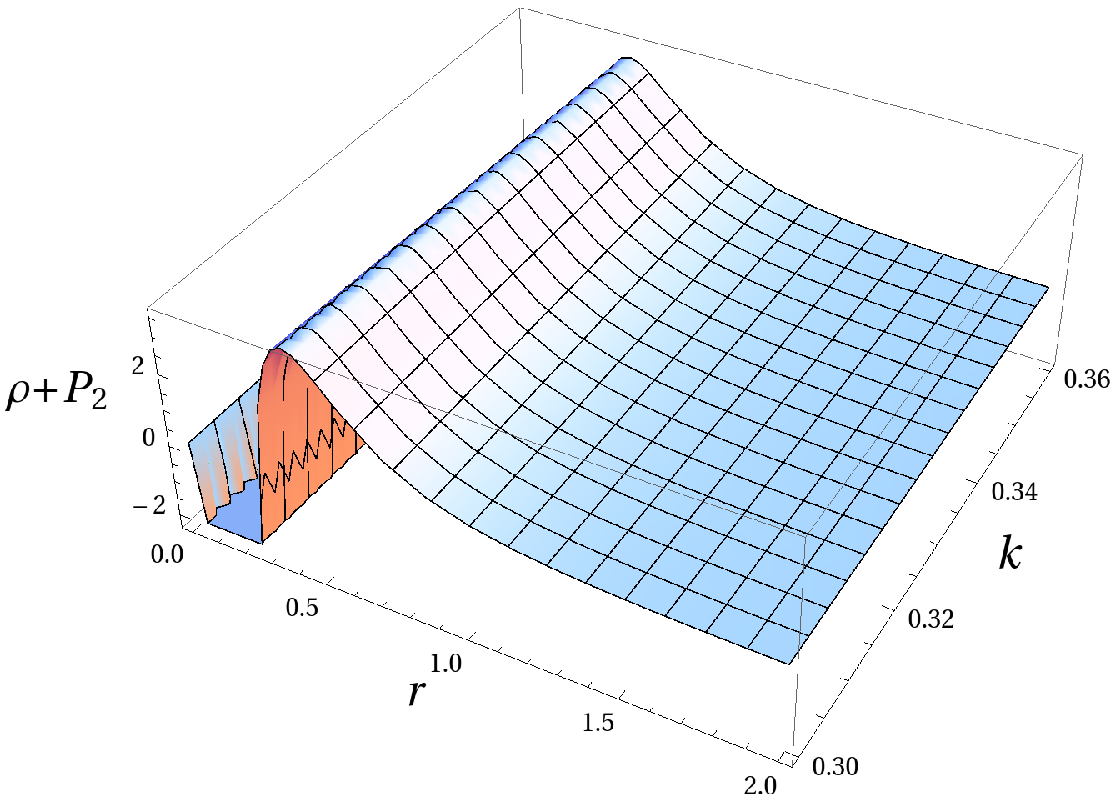}  \\
\includegraphics[scale=0.7]{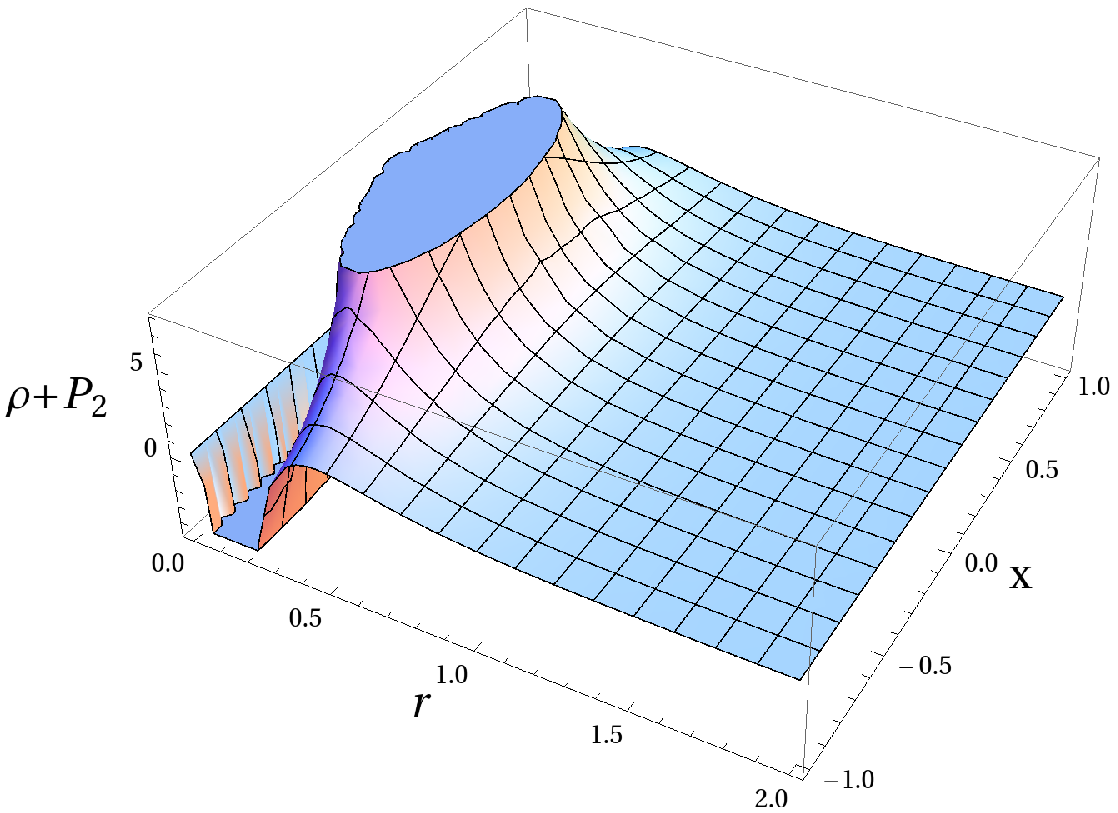} &
\includegraphics[scale=0.7]{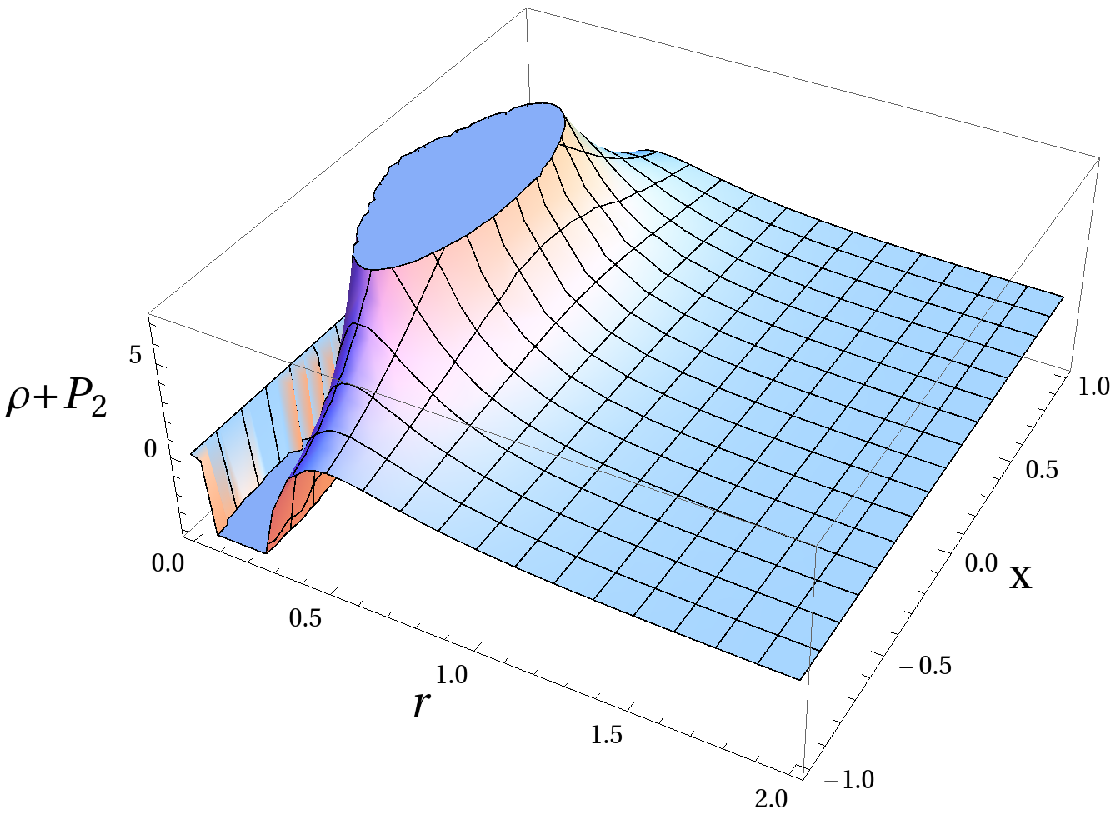}
\end{tabular}
\caption{Plots of $\rho+P_1=\rho+P_2$ vs radius (a) top for the
different values of parameter $k$ ($\theta=\pi/3$ (left) $\pi/4$ (right) ) and (b) bottom for different values of $x= \cos \theta $ ($k=0.3$ (left) $k=0.4$ (right) ) }; \label{rppp}
\end{figure*}

 The weak energy condition requires $\rho\geq0$ and $\rho+P_i\geq0$ ($i=1,\;2,\;3$) \cite{he}.  In fact, for our case one has
 \begin{equation}
 \rho+P_2 = \rho+P_3= -\frac{M k e^{-k/r}\left(k \Sigma - 4 r^3\right) }{r^3 \Sigma^2}
 \end{equation}
 which shows the violation of weak energy condition for a regular  black hole may not be prevented.  The weak energy condition is not really
 satisfied, but the violation can be very small, depending on the
 value of $k$ , as shown in Figs. \ref{rho} and \ref{rppp}.

Next, we add radiation  by rewriting the static solution
(\ref{rotbhtr}) in  terms of the Eddington-Finkelstein coordinates
$(v,r,\theta,\phi)$ \cite{Chandrasekhar}:
\begin{equation}
  v = t+\int \frac{r^2+a^2}{\Delta}, \;
  \bar{\phi} = \phi + \int \frac{a}{\Delta},
\end{equation}
and allow mass $M$ and parameter $k$ to be function time $v$, and
dropping bar, we get
\begin{eqnarray}\label{rotbhvr}
ds^2 & = & -\left( 1- \frac{2M(v)r e^{-k(v)/r}}{\Sigma} \right) dv^2
+ 2 dv dr + \Sigma d \theta^2  \nonumber \\ & - & \frac{4aM(v)r
e^{-k(v)/r}}{\Sigma} \sin^2 \theta dv  d\phi -2 a \sin^2 \theta dr
d\phi  \nonumber \\ & + & \left[r^2+ a^2 +
 \frac{2M(v) r a^2 e^{-k(v)/r}}{\Sigma} \sin^2 \theta
\right] \sin^2 \theta d\phi^2.
\end{eqnarray}
Again relating $M(v),\;q(v)$ with $k(v)$ as done earlier, then all
 stresses of the solution (\ref{rotbhvr}) have same form as that of the solution (\ref{rotbhtr}), but
(\ref{rotbhvr}) has some additional stresses corresponding to the
energy-momentum tensor of ingoing null radiation \cite{Carmeli}.
The solution (\ref{rotbhvr}) describe exterior of radiating objects,
recovering Carmeli solution (or rotating Vaidya solutions)
\cite{Carmeli}  for $k=0$, and Vaidya solution \cite{pc} when
$k=a=0$. The radiating rotating solution (\ref{rotbhvr}) is a
natural generalization of the stationary rotating solution
(\ref{rotbhtr}), but it is Petrov type-II with a twisting, shear
free, null congruence the same as for stationary rotating solution,
which is of Petrov type $ D $. Thus, the radiating solution
(\ref{rotbhvr}) bear the same relation to stationary solution
(\ref{rotbhtr}) as does the Vaidya solution to the Schwarzschild
solution.

  To construct the said rotating regular black hole, we have used an exponential regularization factor suggested by Brown \cite{Brown} used in  a quantum gravity that is finite to plank scale.   The mass
functions $m(r)$ is also inspired by continuous probability distributions to replace mass $M$ of the Kerr black hole.

We have given an example of a rotating regular solution that
(\ref{rotbhtr}) contains the  Kerr metric as the special case when
 the deviation parameter, $ k =0$, also for $r \gg k$  behaves as Kerr-Newman
and it stationary, axisymmetric, asymptotically flat. It turns out that the rotating regular black hole
metrics (\ref{rotbhtr}) can  be also obtained via widely used Newman-Janis algorithm \cite{nja}. It will be useful to  further study the geometrical properties, causal
structures and thermodynamics of this black hole solution, which is
being investigated.

 \begin{acknowledgements} We would like to thank SERB-DST
 Research Project Grant NO SB/S2/HEP-008/2014, to thank Pankaj Sheoran and M. Amir for help in
plots and also to Dawood Kothawala for fruitful discussion.
\end{acknowledgements}

\noindent
\end{document}